\documentclass{article}

\usepackage{graphicx} 

\usepackage{caption}

\usepackage[backend=bibtex, style=numeric]{biblatex}

\title{\textbf{Ephemeral Rollups are All you Need}}
\author{Gabriele Picco, Andrea Fortugno}
\date{www.magicblock.gg}

\begin{document}

\maketitle

\begin{abstract} 

In the realm of open and composable gaming, we envision platforms where users actively expand, create, engage, and immerse themselves in a rich world of entertainment. One promising avenue for achieving this vision is through fully on-chain (FOC) games, where both game state and logic reside on the blockchain, maximizing composability. However, we must grapple with inherent limitations and trade-offs, particularly in terms of costs and scalability. 
This paper proposes a framework that leverages the Solana Virtual Machine (SVM) to scale FOC games without state fragmentation or compromised trust assumptions. The framework introduces a systematic approach for discovering, utilizing, and publishing modular pieces of logic as components deeply rooted in the Entity-Component-System (ECS) pattern. To enhance scalability and resource optimization, we introduce the concept of Ephemeral Rollups (ERs) that overcome the tradeoffs of L2 horizontal scaling. These dedicated runtimes can be customized to provide higher operational speed, configurable ticking mechanisms, provable sessions and gasless transactions without composability-scalability tradeoffs. 

\end{abstract}

\section{Introduction}

Fully on-chain (FOC) games intertwine digital experiences with decentralized immutable ledgers. Any participant of a FOC game can integrate additional content, logic or modifications without explicit consent from the original creator ("permissionless modding") or entertain with the creation of "Autonomous Worlds", unstoppable digital Worlds \cite{aw}. Thanks to smart contracts - which enforce the substrate of these virtual realities - it is impossible to unplug them like traditional game servers. Once deployed, a FOC game will run forever. This degree of creativity and persistency introduces a whole new level of significance to games, which become avenues to play, socialize, work, and conduct commerce in an even more pervasive and profound fashion.

In accordance with the definitions given in \cite{aw}, \cite{1kxfoc}, \cite{dojo}, our perspective holds that for a game to be classified as fully-on-chain, it should inherently possess these crucial attributes:

\begin{itemize}

    \item Blockchain as single source of truth: both game state and logic must reside on the blockchain. It serves as a transparent data store, guaranteeing the permanence of all meaningful data and allowing trustless execution of logic. The game is client-agnostic. 
    \item Permissionless Composability: the game can accommodate new extensions, logic and/or components without the need for explicit permission from the original creator. Trustless computation enabled by smart contracts ensures unbounded access for any enhancement ("everything is a public API"). Even though not every plug-in, extension or third-party client will necessarily be used, everyone has a chance to contribute and add their modifications to it. 
    \item Persistency: The game should demonstrate resilience, with no single point of failure. Even though servers or centralized components may serve to augment performances or playability, the game/world should hold the potential to progress without their presence, thereby ensuring the continuous existence of said digital reality.

\end{itemize}

Our conviction is that FOC games will become the key drivers for the inception of innovative platforms and self-reliant ecosystems. These digital worlds will evolve according to the will of consumers and online communities aligned in this creative endeavour. This concept holds the potential to transcend gaming and blur the line between the virtual and the physical world, seamlessly blending the two.

\subsection{Game Logic, State, and Transactions}

In a fully on-chain game, the game logic resides within smart contracts deployed on the blockchain, ensuring transparency and immutability. It comprises the rules, player interactions, and other behaviors that define the gaming experience. \newline

The game state, on the other hand, refers to the real-time status and attributes of the game elements, which are stored on-chain, providing a consistent and synchronized environment for all players.  It reflects the game's current situation, from player positions to assets and scores. \newline

Transactions in this context denote the interactions and exchanges that occur within the game. Players' actions, asset transfers, and other in-game activities are state transitions recorded on-chain, ensuring their integrity and permanent record. The blockchain's decentralization makes these interaction transparent, verifiable, and immutable.

\subsection{Outline}

In the upcoming sections, \ref{s:scalability} and \ref{s:components}, we introduce a performant and scalable SVM-based framework for FOC games and Autonomous Worlds. We dive into some of key features, notably its scalability and the efficient reuse of resources and on-chain components. For an overview of the emerging gaming frameworks and their limitations, refer to Appendix \ref{a:emerging-framework}.  For an overview of the commonly adopted scalability solutions and their limitations and trade-offs, see the Appendix \ref{a:scalability}.

\section{Framework Architecture}
\label{s:scalability}

Developing multiplayer games is a complex endeavour in conventional settings. FOC games further intensify these challenges, as we must adhere to registering every interaction on-chain, a limiting factor both in terms of scalability and costs. These limitations primarily arise from its decentralized nature and the consensus mechanisms employed.

Considering a straightforward scenario, a game with 1000 players, with each player's position updating every 100 ms, results in 10,000 transactions per second. Scaling this to 100,000 players, the rate leaps to 1,000,000 transactions per second — a volume no blockchain, at the time of writing, can support. Even for a high-speed blockchain like Solana, which might manage such traffic in the future, the demands quickly become insurmountable when considering multiple games and higher numbers of players. From a cost perspective, Solana's localized fee market does alleviate some concerns by preventing a single game from affecting fee costs for other games ("noisy neighbor" problem). However, even if the individual transaction fees are low, the cumulative cost of such a vast number of transactions can become untenable for a single fee payer (i.e., the game developer).

\subsection{SVM runtime}

The Scalability in Solana is intrinsically linked to its unique method of state storage and management, distinguishing it from other virtual machines like the EVM. In Solana, everything is an account. Accounts can hold any data, as well as native SOL tokens. 

Broadly, accounts fall into two categories:

\begin{itemize}
    \item Executable (program accounts): These are akin to smart contracts, housing code and often referred to as “programs”
    \item Non-executable (data accounts): These can store tokens or data but lack the capability to execute code.
\end{itemize}

Solana's account model inherently grants programs the privilege to create and oversee specific accounts. This allows developers to create custom rules and logic for their governance. These special accounts are called Program Derived Addresses (PDAs) \cite{pda}. This mechanism empowers programs to endorse various on-chain actions on a PDA's behalf, ensuring the Solana network recognizes and validates these actions without needing a private key \cite{aa}.

Another core feature of the Solana Virtual Machine is that it parallelizes transaction processing.  It identifies which parts of a transaction are non-overlapping and can be processed in parallel, maximizing hardware utilization. Unlike other blockchains that use a merkle tree to keep track of the global state and process transactions sequentially, Solana's Sealevel executes them concurrently, significantly boosting the network's throughput and scalability. 

\subsection{Ephemeral Rollups}
\label{ss:ephemeral-rollup}

The core intuition is that by harnessing the SVM's account structure and its capacity for parallelization, we can split the app/game state into clusters. Users can lock one or multiple accounts to temporarily transfer the state to an auxiliary layer,  which we define as the "ephemeral rollup", a configurable dedicated runtime. This process temporarily allows the sequencer to modify accounts within the ephemeral rollup, with the state being forcibly reverted and unlocked on the L1 if constraints are not met (see section \ref{ss:trust-compute}). Despite this delegation, operations and transactions can still use the delegate accounts as readable on the base layer. Non-delegated accounts remain unaffected and modifiable. The ephemeral rollup operates as a specialized SVM runtime to facilitate high-throughput, low-latency transaction processing. Additionally, this specialized runtime can be customized to include configurations like gasless transactions, quicker block-time, and the inclusion of a ticking mechanism (i.e., an integrated transaction scheduling system like clockwork \cite{clockwork} operated without fees). The entire process is transparent to the end user - a specialized RPC provider can route the transactions to the base layer and the ephemeral rollup(s) in parallel during the game session.
\newline \newline

\begin{figure*}[h!]
    \includegraphics[width=\textwidth]{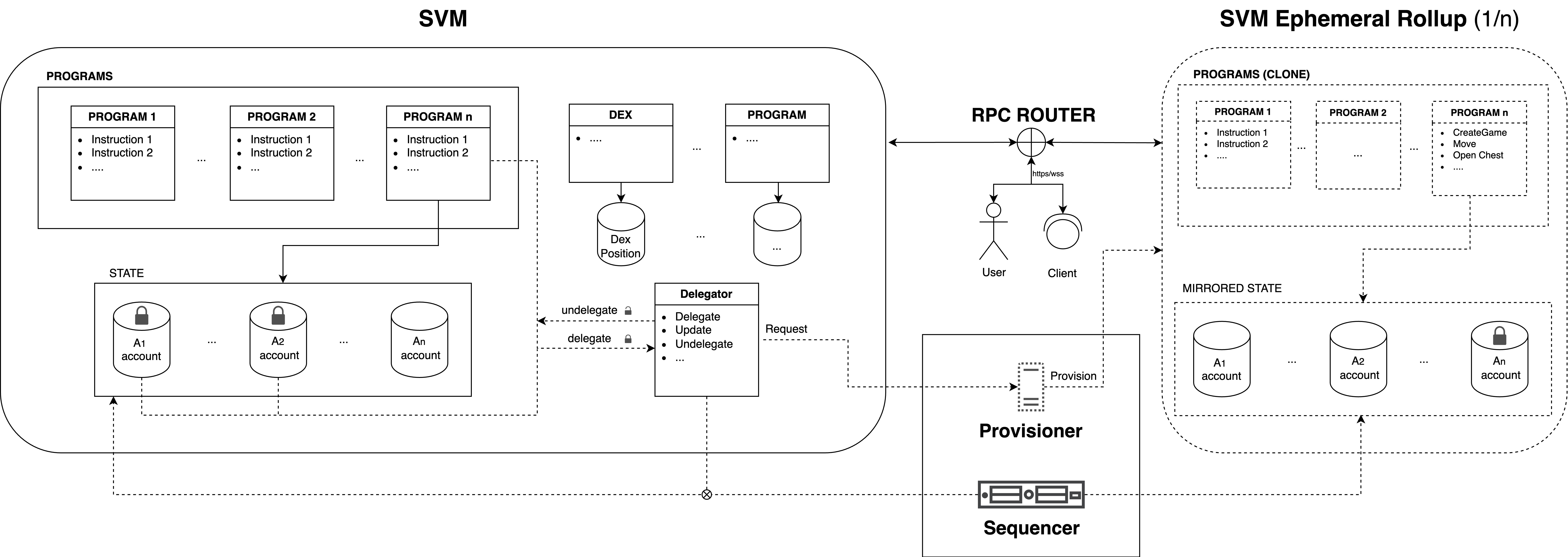}
    \caption{The figure showcases the architecture of the Ephemeral Rollups scalability solution, focusing on how a game program manages its state across multiple accounts. The example illustrates two PDAs, each responsible for recording a state, which could be the position of players. These PDAs are then delegated to the Ephemeral Rollup delegation program (DLP). In the background, the Provisioner constantly observes the DLP program, watching for provisioning requests. Upon detection, it dynamically launches an SVM runtime, adjusting to specific configuration parameters like block time, ticking, and others. Within the Ephemeral Rollup (ER) framework, transactions using these delegated accounts are accelerated. After validation, the account state is updated and finalized on the L1.}
    \label{fig1:arch}
\end{figure*}

Figure \ref{fig1:arch} provides an overview of the system. A program defines executable logic, with its state consisting of various accounts; for a game, PDAs could represent the players' positions, and another account could be a Chest designated for distributing rewards. The program leverages existing programs on the L1, such as an off-the-shelf leaderboard, a contract to mint NFT or an energy system.
\newline
\newline 
The steps involved in provisioning and utilizing the ER are:

\begin{enumerate}
    \item The program communicates with the delegation program (DLP), initiating a transaction that requests an ephemeral rollup provision. This request specifies the ER's configuration details, such as its lifetime, base layer update frequency, targeted transactions per second, and block time.
    
    \item Monitoring the delegation program, the provisioning request, and immediately launches the corresponding runtime based on the configuration.
    
    \item The program delegates accounts to the DLP, granting it the authority for state updating and settlement. From now on, the delegated accounts can only be updated within the ER. This third step can even occur in parallel with, or prior to, the second step.
    
    \item Clients wanting to execute transactions or retrieve data via RPC send their transactions to an RPC router. Depending on the accounts involved in the transactions, this router then forwards the transactions to the appropriate chain, either on the base layer or one of the ERs. A comprehensive discussion of the routing mechanism is available in \ref{ss:routing}.
    
    \item Periodically, or upon ER termination, the sequencer updates the state on the base layer.
    
    \item Once the ER concludes (or if the game program forces closure), account control reverts to the owner program and the provisioner terminates the ER.
\end{enumerate}

To illustrate with a practical example, consider Figure \ref{fig1:arch}. A game program might reward players with an NFT reward upon reaching a specified area on a map. PDAs, tracking players' positions, can be frequently updated within the ER, facilitating a low-latency multiplayer session among players in the ER. When a player reaches the designated map location that triggers a reward, according to the game logic, the client can submit the transaction in the usual manner. If the sequencer doesn’t update rapidly enough, users may first submit an optional state settlement request. Behind the scenes, the Remote Procedure Call (RPC) router relays this transaction to the base layer. Operating on the base layer, the game program can accurately process this transaction and dispense the reward. This is because the position account, which is read-only, reflects the updated location, and modifications can be made to an account that wasn’t delegated through the delegation program.

\subsubsection{Avoiding Fragmentation and Benefits}

The benefit of ER is that programs and assets reside directly on the base layer. Transactions can be accelerated through ERs, which are fully compatible with the Solana Virtual Machine (SVM) down to the bytecode level. Any improvements or advancements at the base layer are immediately available, without the need to modify or re-deploy programs. 

Ephemeral Rollups has the following benefits:

\begin{itemize}

    \item Developers deploy programs to the base layer (e.g., Solana),  rather than on a separate chain as it would normally happen with rollups. Programs reside on the base layer and can interact with any existing protocol and assets. ERs don't fragment the existing ecosystem and allow the speed-up of targeted operations without creating an isolated environment.
    \item Users, developers and programs using ERs can take advantage of Solana's infrastructure. This includes programming languages, code libraries, testing tools, client software, deployment infrastructure etc.
    \item The specialized runtime can accommodate game-specific customizations (e.g., ticking or passive events, typical in games, as opposed to the event-driven runtime of blockchains) without the need for paying gas fees. 
    \item This approach enables a highly scalable system capable of launching rollups on-demand and auto-scaling horizontally to handle millions of transactions without the tradeoffs of traditional L2s.
    
\end{itemize}

Ephemeral Rollups allow FOC games to scale without compromising composability on the base layer.

\subsubsection{Reading and Modifying the state}
\label{ss:routing}

\begin{figure*}
    \centering
    \includegraphics[width=0.5\textwidth]{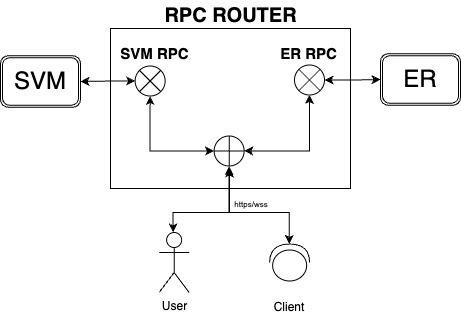}
    \caption{The RPC router module routes both HTTPS and WebSocket requests either to the base layer RPC or to the Ephemeral Rollup RPC, based on the accounts being accessed or utilized.}
    \label{fig2:route}
    
\end{figure*}

As highlighted in the preceding sections, the state is maintained in accounts. These accounts can concurrently exist on the base layer and within an ephemeral rollup (ER). While accounts are universally readable from both layers, they can be modified either from the base layer or the ER. It's important to note that if a user or client seeks to access data from an account, the most recent or updated data might reside in the ER, if it hasn't been settled on the base layer yet. Consider a scenario where a game client employs a WebSockets connection to display other players on a map, and the position PDAs are delegated to the ER. In such a case, a WebSockets connection to the ER would allow observation and streaming of transactions with a latency and block time comparable to a traditional multiplayer game server, typically between 10 to 100 milliseconds.

The added complexity of determining where to submit transactions and where to retrieve data from can be abstracted using an RPC router, as depicted in \ref{fig2:route}. Clients can subscribe and send transactions to the RPC router as they would to a traditional RPC. The RPC Router then directs transactions and connections either to the base layer or to the ER RPC, depending on the accounts involved.
\newline \newline
Considering the following scenarios:
\newline \newline
\textbf{For Reading the state:}

\begin{itemize}
    \item If the request accesses accounts delegated to an ER, the RPC router relays the requests to the ER RPC.
    \item Otherwise, the base layer RPC processes the request.
\end{itemize}

\textbf{For Sending transactions:}

\begin{itemize}
    \item If the transaction contains readable accounts from the base or ER layer and all writable accounts are delegated to the ER, the transaction is relayed to the ER RPC.
    \item If the transaction includes readable accounts from the base or ER layer and no writable accounts are delegated, the transaction is relayed to the base layer RPC.
    \item A challenging scenario arises when a transaction aims to modify accounts present in both the ER and the base layer. There are two options to resolve this situation. The first is to force an intermediate settlement on the base layer, undelegate the accounts and send the transaction through the base layer RPC. The second is to discard the transaction as invalid. The preferred solution depends on the use case. 
\end{itemize}

Additionally, the RPC Router provides the correct blockhash for transactions, a fundamental component in how Solana ensures security and prevents double spending \cite{solana}.

\subsubsection{Account/programs lazy loading and Data Availability}
\label{ss:lazy-load}

An interesting feature to note about ephemeral rollups is the ability to perform lazy loading of the accounts/programs used in transactions. If a transaction uses readable accounts for state or programs (logic) that are part of the main layer, the ER can clone the account state as needed, without having to initialize a complete clone of the main chain. Transactions are available for a specific time window. They can also be indefinitely stored in a compressed account on Solana \cite{compressed-accounts} or within an external DA layer such as Celestia \cite{LazyLedger}. Transactions persist after the ephemeral rollup has been closed, making it easier to verify the correctness of the executed computation.

\subsection{Security and Computation Trustworthiness}
\label{ss:trust-compute}

The two critical security aspects of the ER are as follows:

\begin{itemize}
    \item Preventing an indefinite lock of delegated accounts or compromised state.
    \item Consensus and verification of ER's computation.
\end{itemize}

The mechanism to avoid the first point hinges on the nature of Ephemeral Rollups; as the name suggests, they are ephemeral. Anyone can request the undelegation of accounts when the life of the ER ends by interacting with the program in the base layer. Furthermore, the program that carried out the delegation can revoke it or request settlement at any time, thereby forcing the closure of the ER.

The second point is a common challenge in all rollup architectures, typically addressed through optimistic rollups or ZK rollups. For a detailed description of zk-Rollups, see \cite{ethereum2023zk}; for Optimistic Rollups, see \cite{ethereum2023optimistic}; and for a comparison, see \cite{cryptoslav2021}. \cite{bat} describes ZK channels, a conceptually similar idea for the gaming use-case. As mentioned, transaction history remains available natively in a compressed account on Solana or an external DA layer like Celestia \cite{LazyLedger}). The proposed approach combines these two techniques: it initially performs an optimistic computation for speed, and then produces a ZK proof (e.g., through RISC0 \cite{RISCZero}) for verification. This proof can be efficiently verified on the base layer in a single transaction. 
Anyone can become a Provisioner for the Ephemeral Rollups sessions by staking a predetermined amount of tokens to the ER program, essentially creating a DePIN infrastructure \cite{solana_depin}. 
Furthermore, ER participants themselves can act as SVM diet clients \cite{tiny}, thereby enhancing the trustworthiness of the network.

\section{SVM as a Multiplayer Game Server}

The architecture presented addresses the challenge of processing millions of transactions for an arbitrary number of games. By horizontally scaling the traffic across multiple SVM ERs, it ensures high throughput and low latency, without compromising composability on the base layer, making it ideal for multiplayer gaming scenarios.

Estimating the theoretical throughput of this system is complex due to its dependency on the number of ERs and the sharding strategy. However, a comparative analysis with a multiplayer game server like Nakama \cite{nakama} can provide a performance benchmark. Nakama is renowned for its low-latency performance, often achieving latency close to 50ms, and high-throughput capabilities, supporting thousands of concurrent players in real-time multiplayer games (see \cite{nakama_benchmarks} for a more detailed Nakama benchmark).

As of the time of writing, Solana Sealevel can achieve a throughput ranging from 50,000 to 65,000 TPS \cite{compass-perf}. The advent of newer runtimes, such as Firedancer \cite{firedancer}, is anticipated to further boost the throughput to more than 100k TPS. This level of throughput is already substantial, yet the ER can further enhance the throughput with an arbitrarily large number of dedicated runtimes. In terms of latency, Solana produces blocks approximately every 400 milliseconds. However, the customizable ER runtime can decrease this time to between 10 and 50ms. It's worth noting that the reported performance figures for Solana are based on a network comprising around 2,000 nodes.

Solana's technology, particularly the SVM, could foster a new era of decentralized gaming. The Solana Virtual Machine, with its efficient transaction processing and support for decentralized applications, emerges as a promising infrastructure for hosting multiplayer game servers.

\section{ECS, Mapping and Registry}
\label{s:components}

The framework includes a standardized way to model the game logic, using the ECS (Entity, Components and Systems), which is commonly used in the gaming industry and also adopted in most on-chain engines, such as \cite{dojo} and \cite{mud}.

This pattern decouples logic from state, enabling optimizations in terms of performance and scalability. Additionally, its modular nature facilitates code reusability and extensibility, which are two essential properties for fully on-chain games and autonomous worlds.

The Solana Virtual Machine makes use of a paradigm similar to an ECS. The state (the accounts) and the logic (the programs) are natively separated. This separation is the key mechanism for Solana's parallelism and the scalability solution presented in Section \ref{s:scalability}.

When drawing a comparison with Solana's architecture, we can see how an ECS can be readily implemented:

\begin{itemize}
    \item Entities: An entity is a general-purpose object typically represented by a unique identifier.
    It does not directly contain any data or behavior but serves as an identifier for a collection of components. Entities can be registered within a world instance account on Solana and could also be represented as individual accounts.
    \item Components: Raw data structure that represents some aspect or attribute of an entity. For example, a PositionComponent might just contain x, y, and z coordinates. This concept is similar to how accounts function on Solana.
    \item Systems: Perform the game logic or behavior by acting upon entities with specific components, much like how programs on Solana define the logic and operate on specific accounts.

\end{itemize}

Solana's Program Derived Addresses (PDAs) facilitate the efficient storage and retrieval of components, mimicking the structure of a hash table. \cite{arc} provides an example implementation of an ECS on Solana.

\subsection{Mapping and State listening}

The standardized pattern allows for easy integration of game components with a rendering engine to display the game interface. The presented framework can be viewed as an open and permissionless alternative to a backend intended as a multiplayer game server. For visualization and rendering, existing engines such as Unity \cite{unity3d}, Unreal Engine \cite{unrealengine}, Phaser \cite{phaser}, and others can be used. 
The standardized structure of the components allows for automatic mapping of components and entity properties (abstracting serialization and deserialization, akin to the mechanism of the Anchor framework on Solana \cite{anchor}). State updates can be easily executed and monitored, providing a mechanism similar to the Observer pattern \cite{observer} to listen to state changes and update the rendering.

\subsection{Public Components Registry}

A registry is a complementary infrastructure that works effectively in conjunction with an architecture that leverages the ECS pattern or the general Solana paradigm. A public registry allows for publishing, discovering, and reusing pieces of logic (such as systems and data structures) shared across apps and games. Every program deployed on Solana can be perceived as a system that delineates the logic and receives accounts as input for computation. Specifically for apps and games, we envision a registry housing typical developmental logic - a collision system, an energy system, a movement system etc. — commonly found in traditional engines. Thus, the registry serves as a hub where developers can discover, publish, and utilize systems and components. This facilitates the swift assembly of games by developers without the need to rewrite the entire underlying logic from scratch.

\subsection{Game Loops, Ticking and FRP}

Typically, game engines integrate a Game Loop, which allows for the continuous updating of the state (sometimes referred as ticking) in a loop that performs the following steps:

\begin{itemize}
    \item Process input.
    \item Process game state updates (e.g., physics, AI, object positions).
\end{itemize}

A client-side engine would also include the rendering step. As mentioned in section \ref{ss:ephemeral-rollup}, continuous state updates can be achieved using a transaction scheduling mechanism such as Clockwork. While it is a viable solution, it presents some issues such as the cost of fees and the uncertainty of the exact block/timestamp of the transaction execution. These issues are less prominent in Ephemeral Rollups, where state update transaction costs can be eliminated, and block time can be configured between 1 and 50ms.

While having game loops constitutes the most adopted solution for most game engines, there are also less explored alternatives. There are Haskell engines that instead leverage the functional programming paradigm, such as Yampa \cite{Cheong2005FunctionalPA} and implementations of the functional and asynchronous ECS pattern, such as Apecs \cite{apecs}. The underlying idea is that values are derived through a function instead of evolving the state based on update loops (which can depend on input, time, or other sources). A simple example is an energy system that depends on time, e.g., a function that increases energy by one point every 30 seconds. This paradigm aligns perfectly with the realm of blockchain, where transactions are inherently asynchronous and where the game instance potentially lives forever on the blockchain, evolving over time. This is conceptually similar to Functional Reactive Programming (FRP) and Functional Reactive Animation \cite{elliott1997functional}, which is a programming paradigm that deals with asynchronous data streams and changes in data over time.

Although it is theoretically possible to write an engine that fully utilizes this paradigm, avoiding an update loop, in practice, a combination of the two techniques allows for greater configurability and the ability to use the suitable paradigm considering the game mechanics being developed. 
For example, it is possible to update the position of one non-performing character (NPC) with a loop and derive the position of $n$ other NPCs. The framework will evolve to support both paradigms.

\newpage

\section{Conclusion}

In this article, we have presented the design and architecture of Ephemeral Rollups, a performant and scalable solution for fully on-chain games and Autonomous Worlds. The framework adopts an Entity Component System (ECS) architecture to facilitate the creation and reuse of on-chain components and logic. By leveraging the unique features of the Solana Virtual Machine — its account structure, the delegation of account modifications, and parallelization capabilities — the framework proposes Ephemeral Rollups as a promising path to scale FOC Games. This design enables fast, specialized runtimes that auto-scale horizontally to meet demand without creating isolated environments. As a result, the system maintains composability with existing protocols, tooling and runtime advancements at the base layer without any state fragmentation. 
While this solution is designed for fully on-chain games, it's important to note that the mechanism can be broadly applied to various other use cases.

\printbibliography

\appendix

\clearpage
\newpage

\section{Emerging gaming frameworks}
\label{a:emerging-framework}

Recently, several frameworks have emerged in an attempt to simplify the development of this new domain of games.  Each of these 'on-chain game engines' has distinct features and approaches to facilitate the creation and management of FOC games, each with varying trade-offs. This section delineates the technical nuances and provides a comparative analysis of these frameworks.  

\subsection{EVM frameworks}

One of the earliest on-chain gaming frameworks is MUD \cite{mud}. Lattice, the company behind the MUD framework, was the first to introduce the ECS (Entity-Component-System) paradigm on-chain. While currently on Optimism, MUD aims to resolve the EVM limitation by introducing an opinionated database that indexes the game state from the chain and can be easily queried like any normal relational DB. This enables quicker access to the table of components that constitute an Autonomous World instance compared to retrieving the state directly from the chain. The focus of MUD is on modularity, featuring a system of libraries and plugin-ins. The biggest limitation is the ability to scale or add customization at the runtime level. If MUD scales games using L3, it would fragment the game state across new rollups. 
A similar OP-stack-derived engine is Keystone, developed by Curio, a rollup framework improving the performance of the EVM. Curio is focusing on the more responsive types of games, thanks to the introduction of "ticks", atomic units of time, that ultimately makes the game feel and look more responsive. Similarly to MUD, every game built on Keystone will have to deploy their own independent rollup, with the same fragmentation problem of an isolated state.

LootChain \cite{loot} is yet another low block-time, customizable layer 2 deployed with Caldera. It is developed by the community of Loot and uses ADGL as gas. Aside from a lower block time it doesn't solve the above-mentioned problem of fragmentation across games, nor gas congestion due to the "noisy neighbour" problem.

\subsection{Non-EVM frameworks}

Dojo \cite{dojo} is an open-source framework developed by Cartridge and other contributors on Starknet. The main advantage of Dojo is relying on Starknet prover to create provable-game sessions leveraging zero-knowledge computation and being able to post state diffs rather than entire state updates. The focus on Dojo is on shifting intense computation (i.e., physics) off-chain and proving the integrity of said computation on-chain. Dojo, similarly to other engines, proposes an ECS for the components, tools for scaffolding, indexing and the setup of local nodes to facilitate development. With the release of Madara, a fast L3, Dojo is making explicit that the path to scaling presents the same tradeoffs as other solutions. 

Paima engine \cite{paima} is a stack developed by Paima Studio supporting FOC on Cardano, Polkadot, Algorand and EVM chains. It places emphasis on cross-chain development and at the core it makes use of a state machine that replays the smart contract logic and pushes the update on a sovereign rollup. Paima is explicitly trading off composability in order to achieve higher scalability and parallelization in its state machine. Every game is deployed on its own sovereign rollups. Even a trivial global leaderboard across games built on Paima becomes complicated, if even possible, with this approach.

\subsection{Other frameworks}
World engine \cite{world-engine}, developed by Argus Labs, is an attempt to solve performance issues and introduce a ticking mechanic while also keeping composability in mind. Argus acknowledges that atomic composability is not necessarily a prerequisite for games (i.e., no flash-loans like dynamics) and proposes an EVM L2 with sharded execution environments and customizable runtimes. The insight is that gaming runtimes are inherently different from blockchains as they can have passive events (i.e., day and night cycles, energy regeneration...). Passive events are not generated from user input. This architecture is referred to as "loop-driven" runtime as opposed to "event-driven" runtime.
While on paper, this looks like a promising approach, Argus is effectively bootstrapping a new blockchain from scratch. None of the existing infrastructure, tooling or smart contracts works out-of-the-box on non-EVM shards and the Argus ecosystem is not directly befitting from the broader advancements and activity of the base layer. We believe Autonomous Worlds promise to bring a new level of significance to play, which makes them more likely to succeed in an ecosystem where global payments are settled, userbase and liquidity are abundant, protocols are innovating on new primitives etc. Additionally, atomic composablity can be leveraged to improve UX or to unleash novel use cases in non-trivial manners (i.e., in the context of eSports, an on-chain Bot could borrow a powerful item from an NFT marketplace right before a critical hit to complete a high-stakes fight and repay the loan with the prize won or a user could onramp with FIAT, swap to SOL, create an account abstraction wallet and commit to a prize pool in 1 atomic transaction).

\subsection{Considerations}

In their own unique ways, every framework is trying to increase the performance in terms of blocktime and throughput but introducing composability and network effects tradeoffs. Isolated app chains and current horizontal scaling approaches seem short-sighted, as they fragment the state and reduce synergies with other base layer protocols. 
If it's true that Autonomous Worlds introduce a new level of significance in games and enable virtual experiences that act as avenues for commerce, play and work, we believe blockchains settling large volume of real-world economic activity, RWO, consumer-facing applications, decentralized Physical infrastructure, and synchronizing state at low latency globally are a better fit for these kind of new experiences.

\section{Scalability Approaches}
\label{a:scalability}

A prevalent approach to address scalability challenges involves layers 2, protocols built on top of the base blockchain (Layer 1) to increase transaction speed and reduce costs, and partitioning traffic into isolated segments, known as sharding. Both methods share similarities and scales by segregating the state and computation into independent layers/shards. A similar technique is often employed in multiplayer games, where the multiple server instances manage traffic, e.g., sharding by geographic zones, whether virtual or real-world regions. Layer and Shards must be interoperable; for instance, a player might transition from one map area to another, effectively moving between shards. This method of scalability is incorporated in most game engine described in section \ref{a:emerging-framework}(directly or indirectly). This solution, taken to the extreme, can lead to app-specific L3s, where state and traffic are sharded for each app/game, an emerging scaling strategy for games built on the dojo and mud engines.

This technique inherently leads to state fragmentation, complicating the composability of games, applications, or other protocols on the blockchain. While bridges can solve this issue, they've historically been unreliable, added user interface complexity, and introduced potential sources of bugs. Additionally, it's impossible to execute atomic transactions, which are transactions that are fully executed or not at all. Such atomicity can be essential for specific game mechanics, particularly if a game has economic implications or interfaces with atomic-requiring protocols.

Both solutions can use an improved runtime that can be faster and more performant. In this category, we can conceptually identify two typologies:

\begin{itemize}
    \item The first encompasses various runtime implementations that, while diverse, remain compliant with the parent virtual machine's interface. Speed and performance can be superior either due to a more efficient implementation (for example, an optimized implementation written in Rust vs. Python for the EVM). Alternatively, it might result from different underlying assumptions about security and decentralization. Examples include employing a novel or more centralized consensus mechanism or using an EVM runtime capable of creating a zero-knowledge proof of the execution, decoupling computation and consensus/verification.
    \item In contrast, the second category adopts a distinct state machine, free from the limitations of the base layer's virtual machine. This offers an avenue to leverage optimized runtimes—like Paima \cite{paima} and the World Engine \cite{world-engine} architecture (with Go and Nakama as a multiplayer game server). The state machine define it's own state transition mechanism, and eventually also the consensus and settlement on the base layer.
    
\end{itemize}

Both categories fall under the umbrella of rollups, further classified into distinct categories like ZK, optimistic, smart contract rollup, and sovereign, contingent on the mechanisms of execution, settlement, consensus, and data availability they employ. For a more detailed description, refer to section \ref{ss:rollups}.

\subsection{Scalability vs Composability}

The performance and throughput of the execution layer are constrained by the features of the virtual/state machine and its runtime (the implementation). One clear advantage of using a runtime compliant with the specifications of the virtual machine is maintaining compatibility with the base layer. For instance, with the EVM, protocols and tech stacks are typically reusable or deployable without modifications across any EVM-compatible blockchain and L2/L3 solutions. Such compatibility fosters the growth and rapid expansion of an ecosystem. It also facilitates the integration of technological and application innovations from diverse ecosystems, leading to a general improvement in virtual machines. Morover, users and products can easily be acquired and shared between compatible chains. However, the downside lies in architectural limitations, mainly arising from the blockchain's emphasis on security, decentralization, immutability, and consensus.

Conversely, using a custom or pre-existing state machine not tailored for blockchain allows for performance, latency, and throughput optimization, but complicates composability and typically reduces trust and security assumptions. Additionally, protocols and products that rely on this state machine cannot be transferred between ecosystems, leading to increased fragmentation.

Our framework achieves both benefits, minimizing the scalability/composability trade-off. By leveraging the unique features of the Solana Virtual Machine, described in subsequent sections, it can accommodate any runtime for the SVM and scale it horizontally without compromising the ability to interact with protocols/components on the base layer, even atomically.

\end{document}